\documentclass{article}

\usepackage[nonatbib, final]{neurips_2022}

\usepackage[utf8]{inputenc} 
\usepackage[T1]{fontenc}    
\usepackage{hyperref}       
\usepackage{url}            
\usepackage{booktabs}       
\usepackage{amsfonts}       
\usepackage{nicefrac}       
\usepackage{microtype}      
\usepackage{xcolor}         
\usepackage{amsmath}
\usepackage{graphicx}

\title{Applications of Differentiable Physics Simulations in Particle Accelerator Modeling}

\author{%
  Ryan~Roussel \\
  SLAC National Accelerator Laboratory\\
  Menlo Park, CA 94025 \\
  \texttt{rroussel@slac.stanford.edu} \\
  \And
  Auralee Edelen \\
  SLAC National Accelerator Laboratory\\
  Menlo Park, CA 94025 \\
  \texttt{edelen@slac.stanford.edu} \\
}

\begin{document}

\maketitle

\begin{abstract}
  Current physics models used to interpret experimental measurements of particle beams require either simplifying assumptions to be made in order to ensure analytical tractability, or black box optimization methods to perform model based inference. This reduces the quantity and quality of information gained from experimental measurements, in a system where measurements have a limited availability. However differentiable physics modeling, combined with machine learning techniques, can overcome these analysis limitations, enabling accurate, detailed model creation of physical accelerators. Here we examine two applications of differentiable modeling, first to characterize beam responses to accelerator elements exhibiting hysteretic behavior, and second to characterize beam distributions in high dimensional phase spaces.
  
\end{abstract}

\section{Introduction}
Particle accelerators are indispensable tools for making discoveries in the physical, material, chemical and biological sciences.
Detailed physics models have been developed to describe the generation and manipulation of particle beams for these applications.
Despite this wealth of knowledge, accurately replicating realistic conditions in the physical accelerator is challenging due to the limited availability of measurements and the complex nature of real accelerator components and beam distributions.
In this work, we examine the application of differentiable computing techniques combined with machine learning towards enabling detailed, accurate modeling of realistic accelerator elements and beam properties.

\section{Hysteresis Modeling}
Hysteresis is a well-known physical phenomenon where the state of a given system is dependent on its historical path through state-space.
Hysteresis effects in magnetic \cite{sammut_measurement_2008}, mechanical \cite{huque_accelerated_2015} and material \cite{turner_no_2022} elements of particle accelerators makes optimizing the performance of current accelerator facilities used for scientific discovery challenging using standard black box optimization algorithms, such as Bayesian optimization (BO) \cite{snoek_practical_2012}.

\begin{figure*}[ht]
    \includegraphics[width=\linewidth]{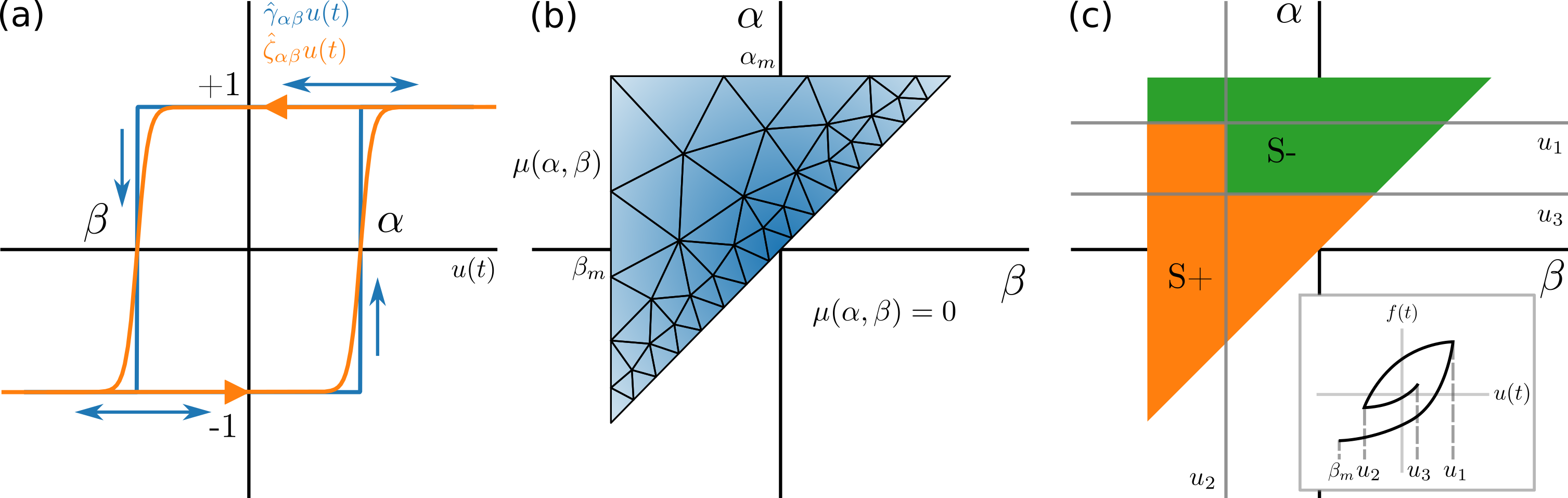}
    \caption{Elements of the differentiable non-parametric Preisach hysteresis model. (a) Output of the hysteron operator $\hat{\gamma}_{\alpha\beta}$ and the approximate differentiable hysteron operator $\hat{\zeta}_{\alpha\beta}$ acting on the input $u(t)$.  (b) Discretization of the density on the Preisach ($\alpha$-$\beta$) plane. Note that $\mu(\alpha,\beta) = 0$ if $\alpha<\beta$, $\alpha > \alpha_m$ or $\beta < \beta_m$ where $\alpha_m,\beta_m$ are equal to the maximum and minimum inputs of the model respectively. (c) $S^+$ and $S^-$ sub-domains after three time steps, where $u_1>u_3>u_2>\beta_m$, assuming that all hysterons are in the negative state initially. Inset: Corresponding model output. }
   \label{fig:preisach_cartoon}
\end{figure*}

The Preisach model \cite{mayergoyz_generalized_1988, bertotti_science_2006} is a numerical representation of hysteresis, comprised of a discrete set of \emph{hysterons}, which when added together, model the output of a hysteretic system $f(t)$ for a time dependent input $u(t)$.
Given a set of discrete time ordered inputs $u_i = u(t_i)$, the hysteron state is represented by the hysteron operator $\hat{\gamma}_{\alpha\beta}$ shown in Fig. \ref{fig:preisach_cartoon}a, which has an output of $\pm1$, where $\alpha$ and $\beta$ describe the input required to switch the hysteron between its two possible states. 
The number of hysterons with values ($\alpha,\beta$) is given by the hysteron density function $\mu(\alpha,\beta)$, plotted on the Preisach ($\alpha$\nobreakdash-$\beta$) plane (Fig. \ref{fig:preisach_cartoon}b).

The Preisach model output is represented by
\begin{equation}
    f(t) = \hat{\Gamma}u(t) = \iint_{\alpha\geq\beta}\mu(\alpha, \beta)\hat{\gamma}_{\alpha\beta}u(t) d\alpha d\beta
    \label{eqn:preisach_model}
\end{equation}
where $\alpha\geq\beta$ results from physical conditions of the hysteron operator.
This integral is evaluated through a geometric interpretation, shown in Fig. \ref{fig:preisach_cartoon}(c).
Given the sequence of input values $u_i$, we can determine sub-regions of the Preisach plane, $S^+$ and $S^-$, where hysteron operators output positive and negative states respectively using a geometric interpretation.

Fitting a Preisach model to experimental data requires the determination of the hysteron density function $\mu(\alpha, \beta)$, often referred to as the \emph{identification problem}. 
We discretize the hysterion density function onto a fine mesh $\mu_i = \mu(\alpha_i, \beta_i)$ and treat the density at each mesh point as a free tuning parameter to fit experimental measurements.
Due to the large number of mesh points and corresponding free parameters, gradient enabled optimization must be used to fit the model to experimental measurements.
This is achieved by implementing the Preisach hysteresis calculation in PyTorch \cite{paszke_pytorch_2019}, which uses backwards auto-differentiation to efficiently calculate gradients.

Our goal is to determine hysteresis properties from beam-based measurements, so we combine the differentiable hysteresis model with a Gaussian process (GP) model \cite{rasmussen_bayesian_2006}, implemented in GPyTorch \cite{gardner_gpytorch_2018}, to describe beam propagation as a function of magnetic fields due to hysteresis. 
We simultaneously infer hysteresis model parameters and GP hyperparameters from training data by maximizing a differentiable calculation of the log likelihood with respect to the parameters of the joint model through gradient descent. 

\begin{figure}[h]
    \includegraphics[width=\linewidth]{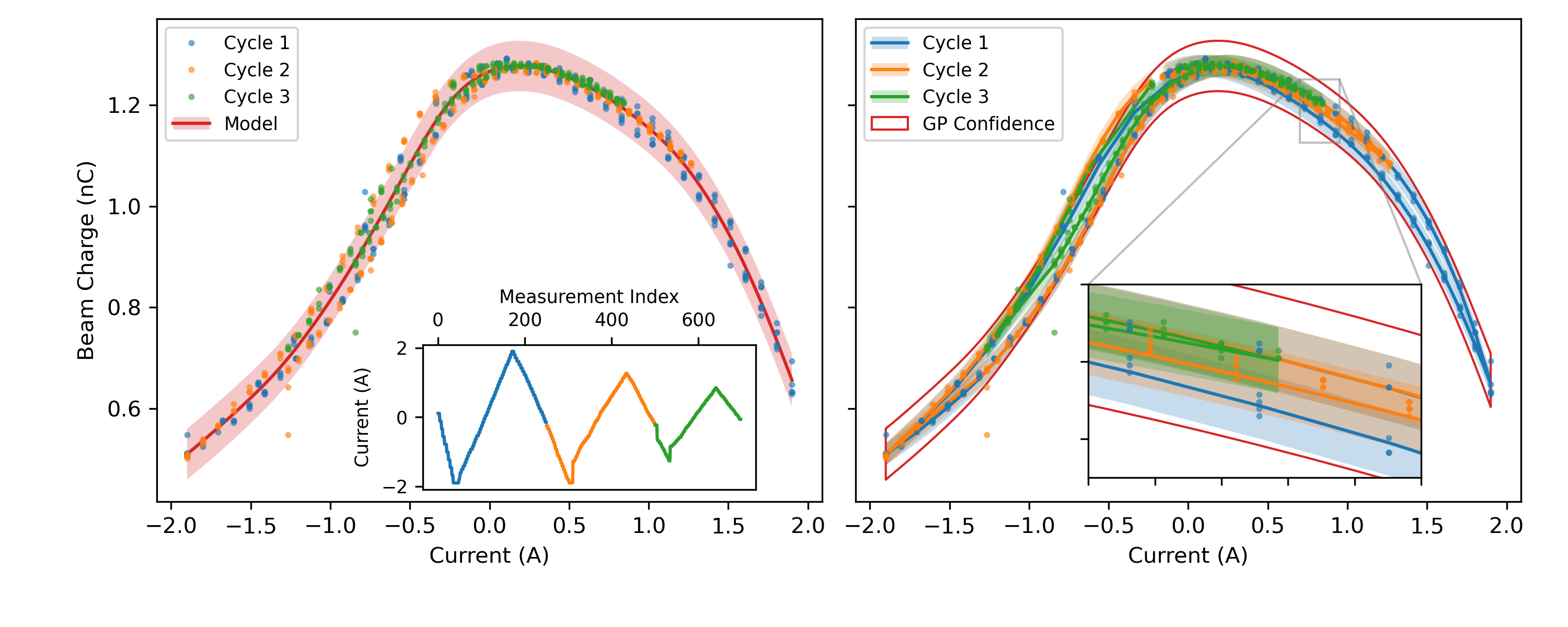}
    \caption{Comparison between GP modeling and joint hysteresis-GP modeling of beam transmission as a function of quadrupole current at the APS injector. (Left) GP model prediction with training data over three cycles (see inset). Shading denotes 2$\sigma$ confidence region. (Right) Hysteresis-GP model prediction, colored by cycle index, compared to confidence interval of normal GP model.
    }
   \label{fig:aps_hybrid_fitting}
\end{figure}

We demonstrate the effectiveness of our joint hysteresis-GP (H-GP) model by fitting the beam response with respect to the current applied to a focusing magnet located in the Advanced Photon Source (APS) injector \cite{sun_recent_2021}.
Measurements from this experiment, shown in Fig. \ref{fig:aps_hybrid_fitting}, have two sources of uncertainty, one from random noise inherent in the accelerator (aleatoric uncertainty) and one due to the unknown properties of magnetic hysteresis (epistemic uncertainty).
We trained two models on the entire data set using L-BFGS with gradient information, the results of which are shown in Fig. \ref{fig:aps_hybrid_fitting}.
First, we trained a normal GP model (Fig. \ref{fig:aps_hybrid_fitting}a) which does not take into account the existence of hysteresis.
As a result it interprets epistemic errors due to hysteresis as aleatoric uncertainty, overestimating uncertainties in portions of the input domain. 
However, the joint hysteresis-GP model (Fig. \ref{fig:aps_hybrid_fitting}b), is able to resolve hysteresis cycles inside the data, removing epistemic uncertainties in the model prediction, thus improving model accuracy and reducing uncertainty.
The increase in accuracy from joint hysteresis-GP models has ramifications for model-based, online optimization of accelerators using BO \cite{roussel_differentiable_2022}.

\section{Phase Space Reconstruction}
Tomographic measurement techniques are used in accelerators to determine the density distribution of beam particles in phase space $\rho(x, p_x, y, p_y, z, p_z)$ from limited measurements \cite{McKee_phase_1995,Hancock_tomographic_1999,Stratakis_phase_2007,yakimenko_electron_2003,rohrs_time-resolved_2009, gordon2022four}. 
While these methods have been shown to effectively reconstruct 2D phase spaces from image projections using algebraic methods, application to higher-dimensional spaces requires independence assumptions between the phase spaces of principal coordinate axes $(x, y, z)$, complicated phase space rotation procedures \cite{hock2013tomographic}, or measurement of multiple 2D sub-spaces with specialized diagnostic hardware \cite{wong_4d_2022}.
It is also possible to directly fit a beam distribution to experimental data through the use of black box optimization algorithms  \cite{wang_four-dimensional_2019, hermann_electron_2021,scheinker2021adaptive} and particle tracking simulations. This however increases the computational cost of inferring the beam distribution from simulation, resulting in simplified beam representations to keep costs within a fixed computational budget.

We propose reducing the computational cost of determining beam phase space distributions by combining methods for parameterizing arbitrary beam probability distributions in 6D phase space using neural networks with differentiable particle tracking simulations.
Particle tracking simulations propagate samples from the initial beam distribution down a model of the beamline to simulated diagnostics.
This allows us to learn the beam distribution from arbitrary downstream accelerator measurements \cite{roussel2022phase}.
Our beam distribution parameterization is heavily inspired by normalizing flows \cite{kobyzev2019normalizing}, however we are only concerned with forward transformations of drawn samples from a base distribution, and not the transformed probability distribution itself (which requires invertable transformations).
Thus we use simple densely connected neural networks instead of a series of invertable transformations.

\begin{figure*}[ht]
    \includegraphics[width=\linewidth]{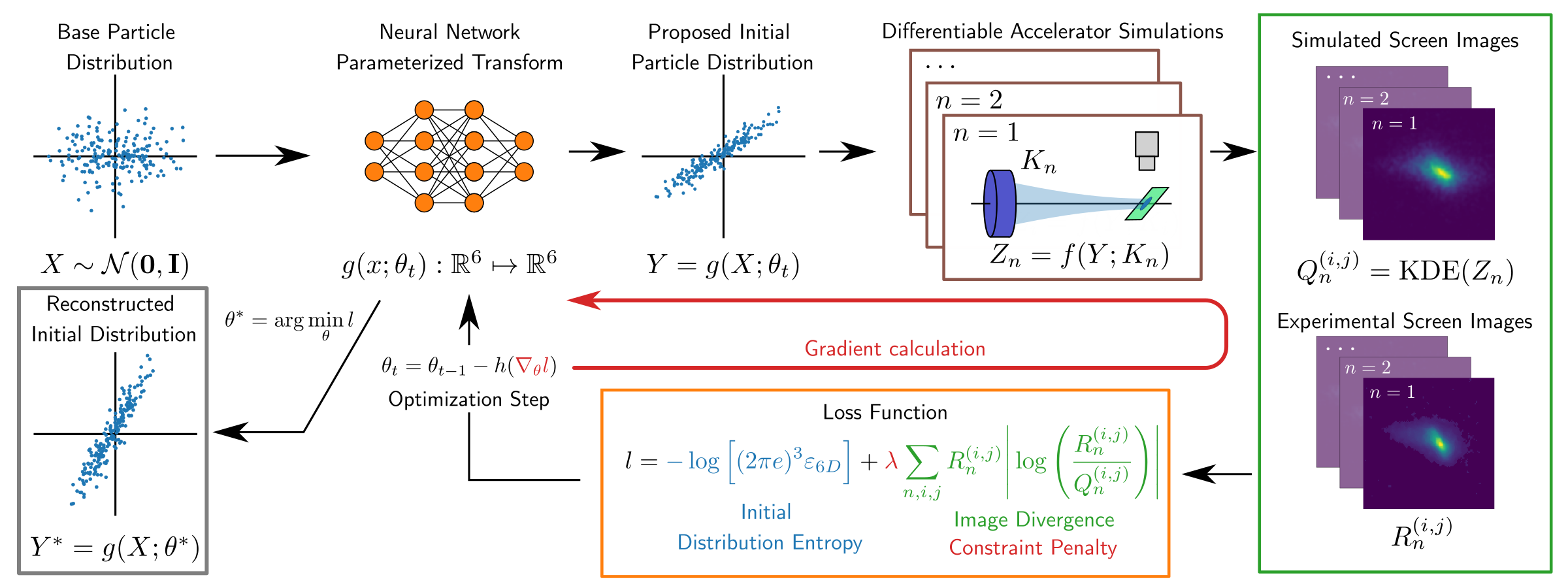}
    \caption{Description of our approach for reconstructing 6D phase space beam distributions.
    First, a base distribution is transformed via a neural network, parameterized by $\theta$, into a proposed initial distribution. This distribution is then transported through a differentiable accelerator simulation of the tomographic beamline. The quadrupole is scanned to produce a series of images on the screen, both in simulation and on the operating accelerator. The images produced both from the simulation $Q^{(i,j)}_n$ and the accelerator $R^{(i,j)}_n$ are then compared with a custom loss function, which when optimized, maximizes the entropy of the proposal distribution constrained on accurately reproducing experimental measurements.}
   \label{fig:reconstruction_cartoon}
\end{figure*}

Fitting neural network parameters to experimental measurements is done by minimizing a loss function to determine the most likely initial beam distribution, subject to the constraint that it reproduces experimental measurements, similar to the maximum entropy tomography algorithm \cite{Hock_A_2013}.
The likelihood of an initial beam distribution in phase space is maximized by maximizing the distribution entropy, which is proportional to the log of the 6D beam emittance $\varepsilon_{6D}$ \cite{lawson1973emittance}.
Thus, we specify a loss function that minimizes the negative entropy of the proposal beam distribution,  penalized by the degree to which the proposal distribution reproduces measurements of the transverse beam distribution at the screen location.
To evaluate the penalty for a given proposal distribution, we track the proposal distribution through a batch of accelerator simulations that mimic experimental conditions to generate a set of simulated images $Q^{(i,j)}_n$ to compare with experimental measurements. 

\begin{figure}[h]
    \includegraphics[width=\linewidth]{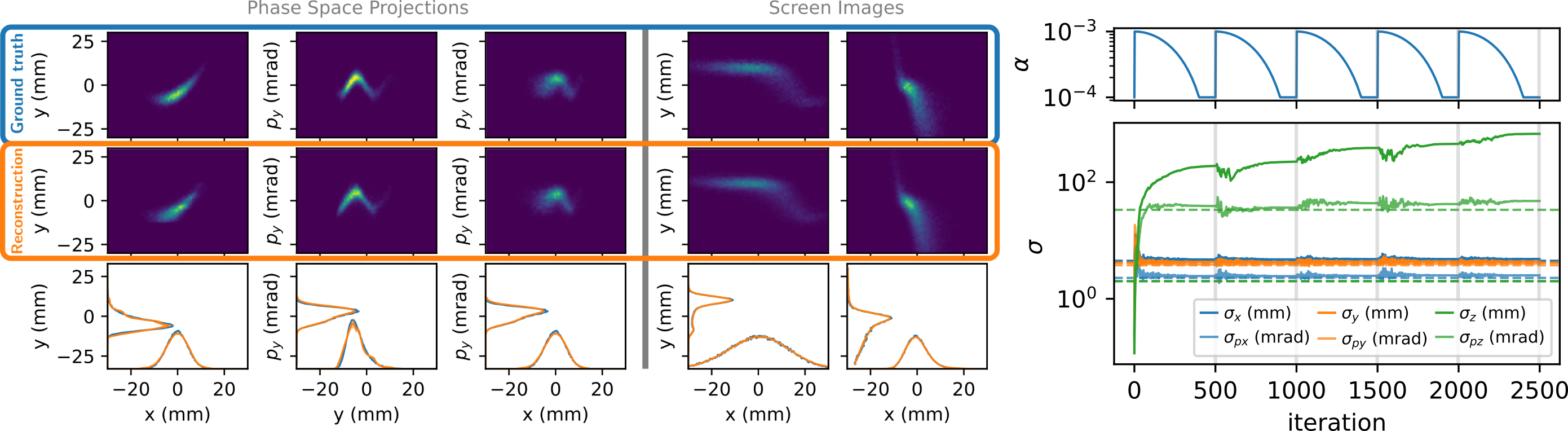}
    \caption{(Left) Comparisons between the synthetic and reconstructed beam probability distributions using our method. A synthetic beam with a tailored phase space distribution is transported through a simulated diagnostic beamline with varying quadrupole strengths and reconstructed using our algorithm. (Right, top) Learning rate schedule for snapshot ensembling. (Right, bottom) Evolution of the proposal distribution beam size in each phase space coordinate during training in synthetic case. Dotted lines denote ground truth values. Vertical lines denote snapshot locations.
    }
   \label{fig:combined_result}
\end{figure}

As in the previous example, a loss function that is \emph{differentiable} is needed to fit the large number of neural network parameters.
Unfortunately, no particle tracking codes which are differentiable with respect to particle coordinates currently exist.
To satisfy this requirement, we implement particle tracking and image creation using \emph{PyTorch} \cite{paszke_pytorch_2019}.
We demonstrated reconstruction of a beam in phase space using a synthetic test case, shown in Fig.~\ref{fig:combined_result}.
We see excellent agreement between the average reconstructed and synthetic projections in both transverse correlated and uncorrelated phase spaces, as well as transverse beam images.

We characterize the confidence of our reconstruction using snapshot ensembling \cite{huang2017snapshot}.
This encourages the optimizer to find multiple possible solutions (if they exist).
It is instructive to examine the evolution of the proposal distribution during model training (shown in Fig.~\ref{fig:combined_result}).
The phase space components that have the strongest impact on beam transport through the beamline as a function of quadrupole strength converge quickly, whereas the ones that have little-to-no impact (e.g. the longitudinal distribution characteristics) do not. 
In particular, we observe that each snapshot converges to beam distributions which have slightly varying energy spreads, signifying uncertainty in that aspect of the beam's phase space, likely due to weak coupling between energy spread and transverse beam size through chromatic aberrations in the quadrupole.
The convergence of the proposal distribution thus provides a useful indicator of which components of the phase space can be reliably reconstructed for arbitrary (possibly unknown) tomographic beamlines.

\section{Conclusion}
Here we have shown the advances in accelerator modeling that can be achieved with differentiable computations.
When coupled with machine learning techniques these models can surpass limitations faced by historically used methods for analyzing experimental data.
Further investments in creating fully differentiable accelerator physics simulations will continue to improve model detail and accuracy.

\section{Impact Statement}
This work is expected to have significant impact on the ability of accelerator diagnostics in the future. As the work is limited within the scope of accelerator science we expect there to be no ethical aspects or future societal impacts beyond the improvement of accelerator operations. 

\bibliographystyle{ieeetr}
\bibliography{my_bib}

\end{document}